\documentclass[sigconf]{acmart}
\usepackage{multirow}
\usepackage{tcolorbox}
\usepackage[T1]{fontenc}
\usepackage{subcaption}
\usepackage{array}
\usepackage[ruled,linesnumbered,vlined]{algorithm2e}

\newcounter{finding}
\newcommand{\parhead}[1]{\noindent\textbf{#1}}

\newcommand{\eat}[1]{\ignorespaces}

\AtBeginDocument{%
  }

\setcopyright{acmlicensed}
\copyrightyear{2018}
\acmYear{2018}
\acmDOI{XXXXXXX.XXXXXXX}
\acmConference[Conference acronym 'XX]{Make sure to enter the correct
  conference title from your rights confirmation email}{June 03--05,
  2018}{Woodstock, NY}

\acmISBN{978-1-4503-XXXX-X/2018/06}

\begin{document}
\title[A First Look at Ethereum Blob Revolution: Market, Strategies, and Optimality]%
{A First Look at Ethereum Blob Revolution:\\Market, Strategies, and Optimality}

\author{Yue HUANG}
\email{yhuang797@connect.hkust-gz.edu.cn}
\affiliation{%
  \institution{The Hong Kong University of Science and Technology (Guangzhou)}
  \city{Guangzhou}
  \country{China}
}
\author{Shuzheng Wang}
\email{swang032@connect.hkust-gz.edu.cn}
\affiliation{%
  \institution{The Hong Kong University of Science and Technology (Guangzhou)}
  \city{Guangzhou}
  \country{China}
}
\author{Yuming Huang}
\email{huangyuming@u.nus.edu}
\affiliation{%
  \institution{National University of Singapore}
  \country{Singapore}
}

\author{Gareth Tyson}
\email{gtyson@ust.hk}
\affiliation{%
  \institution{The Hong Kong University of Science and Technology (Guangzhou)}
  \city{Guangzhou}
  \country{China}
}

\author{Huayi Duan}
\email{huayiduan@hkust-gz.edu.cn}
\affiliation{%
  \institution{The Hong Kong University of Science and Technology (Guangzhou)}
  \city{Guangzhou}
  \country{China}
}

\author{Jing TANG}
\authornote{Corresponding author: Jing TANG.}
\email{jingtang@ust.hk}
\affiliation{%
  \institution{The Hong Kong University of Science and Technology (Guangzhou)}
  \city{Guangzhou}
  \country{China}
}

\begin{abstract}
As a key enabler of Web3, Ethereum has long faced scalability challenges. The recent EIP-4844 upgrade aims to alleviate the scalability issue by introducing the ``blob'', a new data structure for Layer-2 rollups that enables off-chain storage with much reduced costs. Yet, this new mechanism's impact on Ethereum, and the wider Web3 ecosystem, remains largely underexplored. In this paper, we conduct the first large-scale empirical analysis of the post-EIP-4844 ecosystem, leveraging a dataset of $319.5$ million transactions, out of which $1.3$ million are blob-carrying. Our analysis reveals two major trends: (1) average block size has increased $2.5\times$, from $150$ KB to $400$ KB, while the share of conventional transactions has shrunk from over $150$ KB to around $80$ KB; (2) rollups are rapidly migrating from expensive calldata, falling from approximately $7{,}500$ to nearly zero, toward cheap blobs, rising from zero to about $10{,}000$. These shifts introduce a new economic game between block builders and rollups. Thus, we develop a game-theoretic model to characterize their equilibrium strategies: a profit-maximizing inclusion rule for builders, and a cost-minimizing blob batching strategy for rollups. Empirically, however, we find notable economic inefficiencies: for example, $29.48\%$ of blob-containing blocks are built sub-optimally, yielding less revenue than available alternatives. These findings highlight the intricacies of the blob marketplace, and our work has established both methodological and empirical foundations to understand the evolving post-EIP4844 Ethereum ecosystem.

\end{abstract}



\received{20 February 2007}
\received[revised]{12 March 2009}
\received[accepted]{5 June 2009}

\maketitle

\section{Introduction}
\label{sec:intro}
As a major facilitator of Web3, Ethereum is an open platform that facilitates the execution of smart contracts and decentralized applications~\cite{gilbert2022crypto, wang2022exploring}. 
However, as the Ethereum user base has grown, the mainnet has faced significant network congestion, leading to limited throughput, high delays, and prohibitive expense. 

In response, the community has proposed various solutions, among which \emph{Layer-2} protocols are believed to be the most promising. 
Recently, rollups have emerged as the most widely adopted Layer-2 solution, driven by their ability to balance scalability, security, and decentralization. In this paper, we view rollup applications in Ethereum as distinct \emph{rollup} entity. They enhance efficiency by bundling hundreds of transactions into a single batch on a separate chain, and then submitting a concise summary of this batch to the Ethereum mainnet for final settlement. By offloading the bulk of computation, rollups drastically reduce the network's congestion and lower costs for users, all while inheriting the robust security guarantees of the mainnet.

The growing adoption of rollups places significant economic pressure on their operators. To ensure data availability and security, these protocols are required to post compressed transaction data to the expensive ``Calldata'' space on the Ethereum mainnet --- used for verifying data in Layer-2 transactions. They are costly because each byte of calldata consumes on-chain storage and gas, and such data is much larger than typical transactions. 
As user activity has intensified, the frequency and cost of these ``Calldata'' submissions has increased commensurately, creating a major operational burden. At its peak, this ``data tax'' on Calldata cost major rollups a collective sum exceeding $10,000$ USD per day~\cite{0xrob2025dune}. To alleviate this growing concern, the Ethereum Dencun upgrade (\textbf{EIP-4844}) created a dedicated and more cost-efficient data space for rollups known as ``blobs''. Accessed via a new \textbf{type-3 transaction}, blobs allow rollups to offload data to inexpensive off-chain storage, thereby reducing their operational costs~\cite{dencun2024ethereum, vitalik2022eip4844}. 

This new mechanism alters the decision-making process for both block builders and rollups. Block builders, whose primary task is to construct the most profitable block by sorting transactions~\cite{daian2020flash, heimbach2022sok}, now face a dilemma: 
including large blob-carrying transactions may increase revenue, but this also increases the block's propagation time~\cite{yuan2024eip}, raising the risk of the block being orphaned and the builder losing \emph{all} profit; meanwhile, rollup must compete to have their type-3 transactions included by those (selfish) builders. 
This has led to multifaceted strategies~\cite{crapis2023eip}, where rollups must decide on optimal blob bundling and offer compelling priority fees to incentivize builders to accept risky transactions. The result is a novel and complex marketplace for blob inclusion with diverse player moves~\cite{jason2024blob}. 

This paper presents the first study of the dynamics game introduced by the Ethereum Dencun upgrade (EIP-4844). Our research is driven by the following three research questions:
\begin{itemize}\small
    \item \textbf{RQ1: What are current strategies for block construction and Layer-2 data verification?}
    \item \textbf{RQ2: What are the profit-optimal strategies for block builders and rollup?}
    \item \textbf{RQ3: Have the optimal strategies been adopted in-the-wild by existing market players?}
\end{itemize}

To answer the questions, we construct the first large-scale dataset dedicated to analyzing the interactions between builder's block construction and rollup's Layer-2 data verification behavior. This dataset, consisting of over $319$ million transactions and $1.3$ million type-3 transactions, serves as the bedrock for our investigation~(\autoref{sec:5_data}).\footnote{We will make the dataset publicly available in the published version.} 
We observe that the average block size has grown $2.5\times$ to nearly $400$ KB in from March to August 2024. Simultaneously, the space consumed by traditional transactions without blobs shrinks dramatically---from over $150$ KB to just $80$ KB. In parallel, our analysis shows that rollups are gradually moving their data posting from expensive calldata to more affordable blobs~(\autoref{sec:builder}).

These observations inspire us to further inspect the game played in the blob marketplace. Specifically, we construct a game-theoretic model to capture the complex interplay between builders and rollups. To maximize profit, builder must compare the highest priority fee of type-3 transactions with its floor price to make the decision for transaction inclusion, whereas rollups must decide the optimal number of blobs to bundle into each transaction. 
Through rigorous theoretical analysis, we derive the optimal strategies for both of them to achieve equilibrium in the game~(\autoref{sec:Game Model}). 

Leveraging this model, we further examine whether market players are adopting optimal strategies in the wild.
Our investigation reveals the existence of widespread economic inefficiencies~(\autoref{sec:existing_inefficiency}). Builders frequently accept blob transactions even when their revenue fails to cover the opportunity cost of excluding more profitable transactions available in the mempool. In particular, there are $29.48\%$ of blocks containing blobs that were unprofitable. This inefficiency was particularly stark in March 2025 (the month following the upgrade), where potential profits from the mempool were remarkably 5x greater than the revenue builders gained from blobs. 
The situation for rollups is equally concerning. Our analysis uncovers that the common blob submitting strategies from rollups are far from optimal. Rollups such as Scroll and Starknet, which consistently split their blobs into separate type-3 transactions, generate sub-optimal type-3 transactions accounting for $69.85\%$. By adopting our optimal strategy, rollup can substantially reduce these inefficiencies and improve overall profitability.

Our main contributions are:
\begin{enumerate}
    \item We construct a large-scale dataset to analyze the builder-rollup market in the post-EIP-4844 era~(\autoref{sec:5_data}). We uncover the players' trending behaviors for block construction and calldata-blob transition~(\autoref{sec:builder}).
    
    \item We develop a game-theoretic model to describe the market's dynamics and derive the equilibrium strategies for both players to obtain maximum profits~(\autoref{sec:Game Model})    
    
    \item We conduct a comprehensive empirical study of builder and rollup behavior in practice, revealing that both parties operate sub-optimally and hence they can improve their operations. (\autoref{sec:existing_inefficiency}) 
\end{enumerate}

\section{Background}
\label{sec:background}
In this section, we briefly describe the context in this section, including Layer-2 rollup, block builder and details of EIP-4844.
\begin{figure*}[!bpt]
    \centering
    \includegraphics[width=1\textwidth]{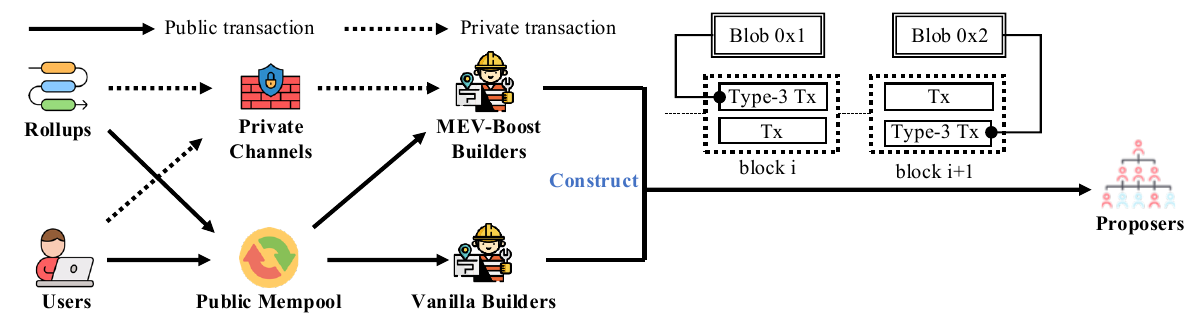}
    \vspace{-24pt}
    \Description{Illustration of the block building process in Ethereum.}
    \caption{Illustration of the block building process in Ethereum. Builder is divided into MEV-Boost builders and Vanilla builders according to their transaction source. Their blocks would contain type-3 transactions introduced by EIP-4844.}
    \label{fig:ethereum_pbs}
\end{figure*}

\subsection{Layer-2 Rollups}
\label{rollup}
Rollups are Layer-2 scaling solutions designed to improve the efficiency and throughput of the Ethereum network by offloading transaction processing from the Ethereum mainnet~\cite{danksharding2024ethereum}. To protect the data availability of Ethereum~\cite{qbzzt2024datastorage}, rollups work by bundling multiple transactions into a single batch. After bundling, the rollup smart contract on Ethereum updates the state with a single, compressed transaction that includes the aggregated data of all individual transactions. This compressed transaction will be submitted to the mainnet, thereby reducing the computational load and gas fee for each individual transaction. Rollups maintain the security and decentralization of Ethereum while providing faster and cheaper transactions, making them a crucial component of the Ethereum scalability roadmap~\cite{corwin2024scaling}.

Rollups come in two main types: optimistic rollups~\cite{ian2024opimistic} and zero-knowledge rollups~\cite{paul2024zk}. Optimistic rollups assume transactions are valid, and only run fraud proofs when discrepancies are detected. Zero-knowledge rollups use cryptographic proofs to validate the correctness of the bundled transactions. 

\subsection{Ethereum Builders}
\label{builder}
Layer-2 rollups periodically publish compressed cryptographic proofs to the Ethereum mainnet, thereby inheriting its security and finalizing transactions. \autoref{fig:ethereum_pbs} illustrates the current architecture of Ethereum. The Proposer Builder Separation (PBS) mechanism splits the role of validators (aka miners) into separate entities: MEV-Boost builders and proposers~\cite{buterin2021proposer, buterin2021state}. MEV-Boost builders are responsible for block construction. They select transactions from the public mempool~\cite{tx2024ethereum} or private channels~\cite{wang2025private}. The constructed blocks are then forwarded to proposers, accompanied by fee commitments offered by the builders. Proposers endorse the block with the highest fee for the finalization of the blockchain. 

Currently, it is optional for Ethereum validators to participate in PBS~\cite{toni2024blobs}. Validators who opt not to adopt the PBS framework and instead build blocks independently are known as Vanilla builders. Vanilla builders receive no private transactions as they are not specialized in block building. They complete both block building and validation tasks simultaneously and currently account for 10\% of all builders in the network~\cite{toni2023mevboostpics}.

\subsection{EIP-4844 \& Type-3 Transaction}
\label{EIP4844}
Rollups use calldata to store state roots and proofs~\cite{qbzzt2024datastorage}, ensuring that the rollup state could be re-established even if a transaction was challenged. Calldata is the payload contained in a transaction that can be used to store arbitrary information on the chain. While calldata is cost-effective for smaller amounts of data, it is permanently stored in Ethereum state and is therefore relatively expensive for larger data requirements. The high cost of calldata storage becomes a bottleneck for scaling.

In order to improve the scalability and data availability of the network, Ethereum proposed a new sharding design: Danksharding~\cite{danksharding2024ethereum}. Danksharding is a unified sharding design that differs from traditional sharding by managing data storage and availability in a more streamlined way. Instead of increasing the space for transactions, Ethereum sharding provides more space for blobs of data, which the Ethereum protocol itself does not interpret. 

EIP-4844, also known as Proto-Danksharding, implements the same transaction formats and validation rules. EIP-4844 was launched as a core part of the Dencun upgrade in 2024~\cite{dencun2024ethereum}, thereby facilitating the network's preparation for full Danksharding by allowing the inclusion of large data in transactions~\cite{qbzzt2024datastorage}.

Much earlier, in 2021, EIP-1559 introduced a new transaction format (type-2) which replaced the legacy type-0 format by using maxFeePerGas and maxPriorityFeePerGas instead of a single gasPrice~\cite{liu2022empirical}.\footnote{Note, type-1 was a transitional format that added an accessList to the legacy model to provide gas discounts for accessing specific contract storage.}
Type-2 transactions always bring a priority fee for inclusion by builders quickly. In contrast, EIP-4844 implements a new type: type-3 transactions~\cite{park2024impact}. These are similar to regular type-2 transactions, but carry an additional data component called a `blob'. A blob is a large, temporary data payload attached to a type-3 transaction. Unlike calldata, blob data is not stored permanently on-chain. Instead, only a commitment to the blob is included on-chain, significantly reducing storage costs~\cite{qbzzt2024datastorage}. These oversized blobs (128 KB) are much cheaper than the equivalent amount of calldata because they are not permanently stored on Ethereum mainnet~\cite{qbzzt2024datastorage}, and rollups can also send their type-3 transactions from public mempool or private channels. This new type of transaction meets the growing need for efficient data processing. 
This change, while lowering the transaction costs, has created a new high-stakes economic game for block builders and rollups. However, the dynamics of this game remain poorly understood. 
\section{Data Collection}
\label{sec:5_data}
We generate our data sets in two stages: (1) data collection and (2) labeling. We cover the period from December 1, 2023 to August 31, 2024. To our knowledge, this is the first time such an extensive label dataset has been constructed.

\begin{figure*}[!bpt]
    \centering
    \begin{subfigure}{1\textwidth}
        \centering
        \includegraphics[width=\linewidth]{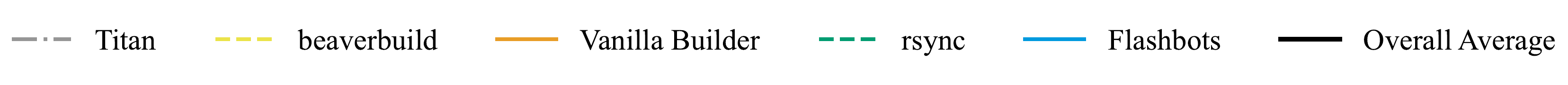}
    \end{subfigure}
    \hfill
    \begin{subfigure}{0.49\textwidth}
        \centering
        \includegraphics[width=\linewidth]{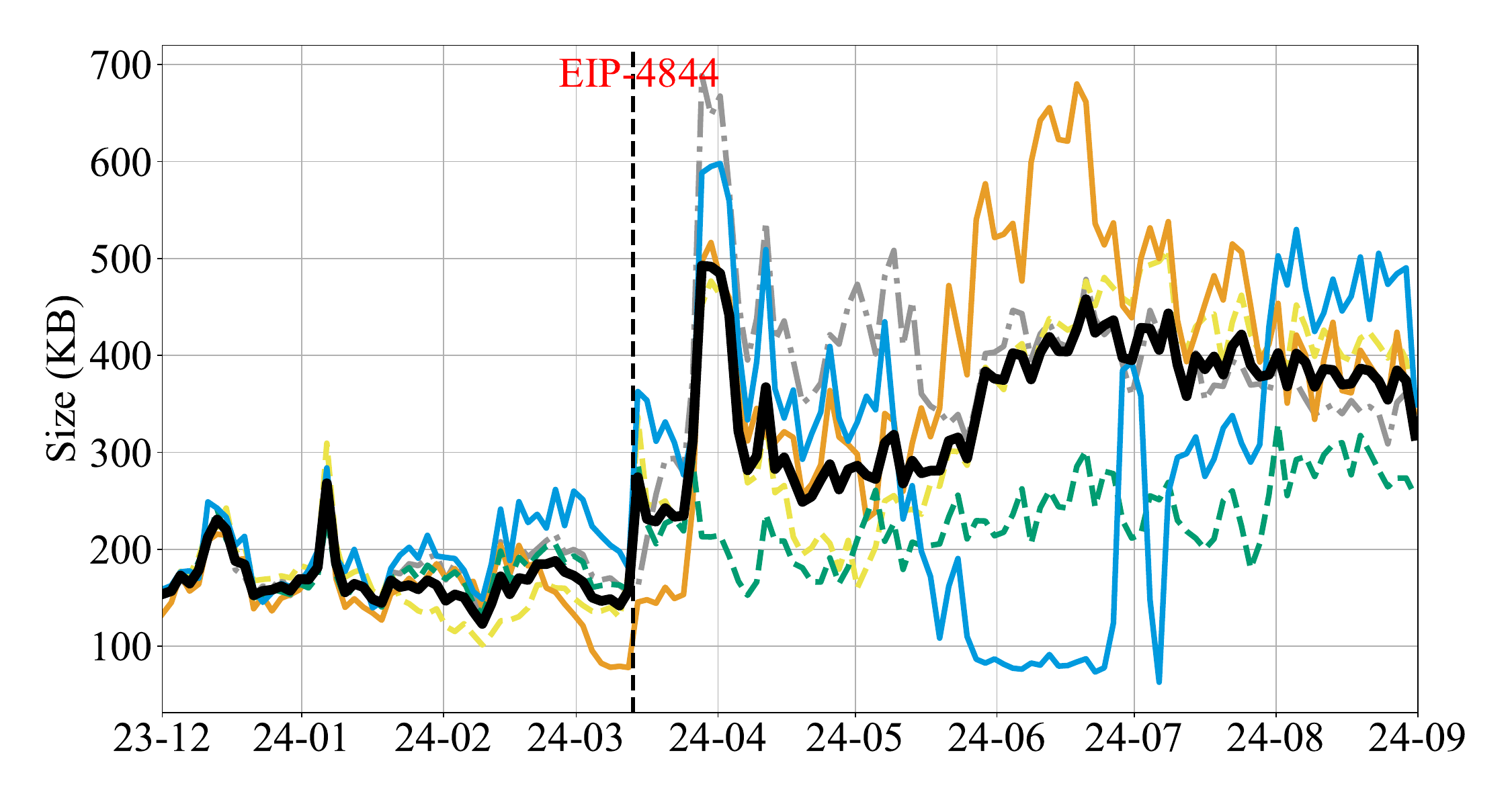}
        \caption{Overall block size.}
        \Description{Block size.}
        \label{fig:avg_block_size_block}
    \end{subfigure}
    \hfill
    \begin{subfigure}{0.49\textwidth}
        \centering
        \includegraphics[width=\linewidth]{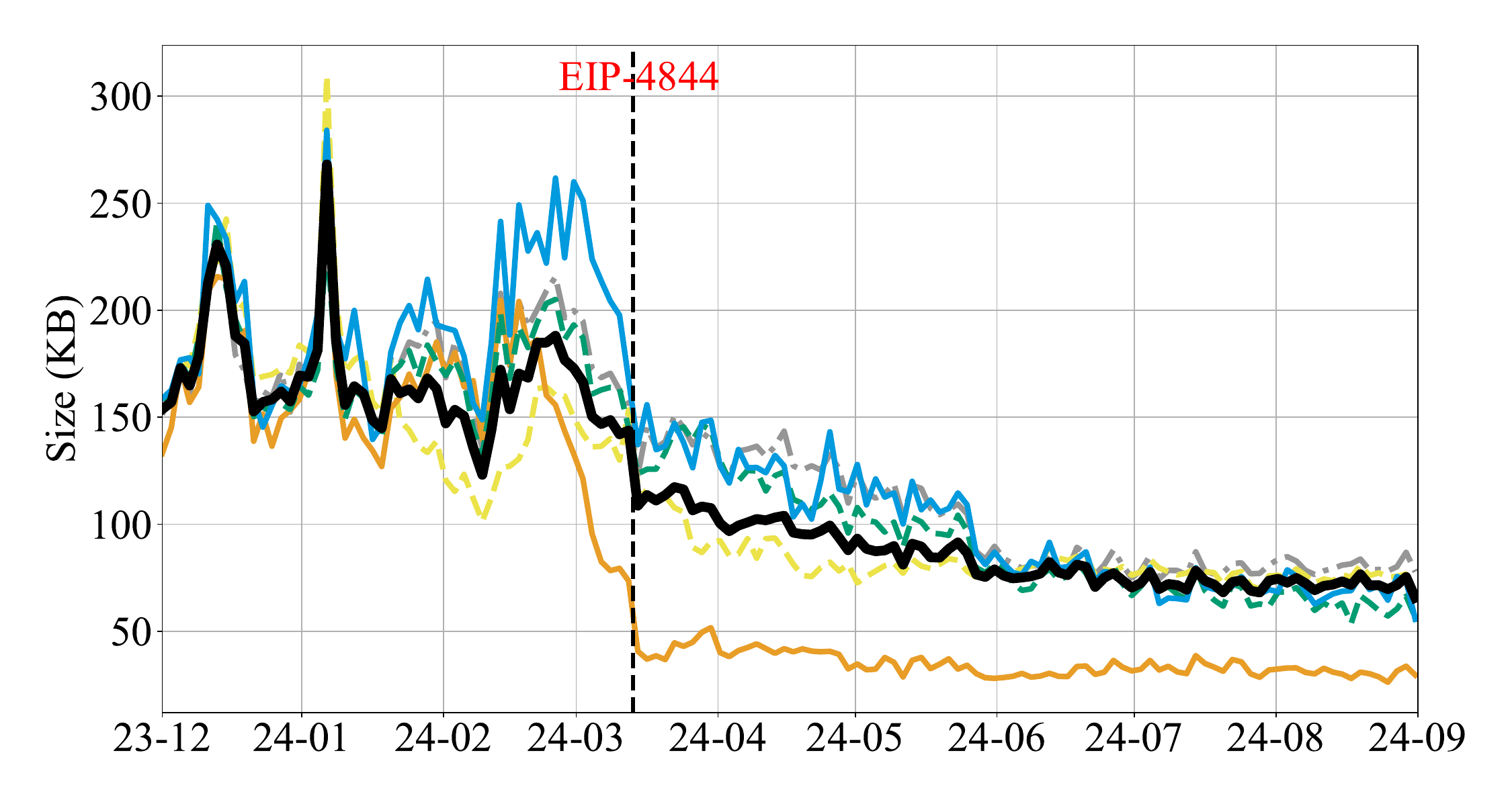}
        \caption{Size excluding type-3 transactions.}
        \Description{Type-3 excluded block size.}
        \label{fig:avg_block_size_non-3}
    \end{subfigure}
    \Description{Average block size from builders.}
    \caption{The average size in block among builders.}
    \label{fig:avg_block_size}
\end{figure*}

\subsection{Data Collection}
\parhead{On-chain Data.~} To build an on-chain dataset, we deploy the Ethereum execution client Erigon~\cite{erigon} and the consensus client Lighthouse~\cite{lighthouse}. The on-chain dataset contains 1{,}967{,}415 blocks, 319{,}529{,}950 transactions and receipts. The dataset also includes the \texttt{hash} address of each blob in 1{,}336{,}822 type-3 transactions.

\label{sec:delay_data_archive}
\parhead{Public Mempool Data.~} 
We next acquire the public mempool archive data covering the same period from Flashbots Mempool Dumpster~\cite{flashbots-mempool}, which is updated every day. The mempool data provides the transaction \texttt{inclusion delay} and the information such as the \texttt{type} and \texttt{size} of a transaction~\cite{chris2023mempoool}. The transaction \texttt{inclusion delay} is calculated from the time interval between the first time the transaction is detected and the time when it is included in a block. The public mempool data range from the same period of on-chain dataset and have 318{,}175{,}240 public transactions. 

\subsection{Labeling}
\label{sec:delay_data_label}
\parhead{Builder Labeling.~} To identify the builder of a particular block, we decode the \texttt{extra\_data}.
To measure our coverage, we compare our builder labeling dataset with existing public dataset~\cite{eden2024dataset} created using builders' \texttt{public keys}. This confirms that our dataset covers over 300 vanilla builders that were ignored in prior datasets.\cite{yang2025decentralization} Our labels contain 56 MEV-Boost builders and 318 Vanilla builders.

\parhead{Rollup Labeling.~} 
To label the observed rollups in Layer-2, we collect the official on-chain address disclosed by~\cite{hildobby2024blobs, victoria2024blob} to match the corresponding rollups. However, we find that not all rollups are correctly identified. To remedy this, we note that some rollups add their chain ID to the \texttt{to} address in their transactions. Thus, to label more rollups, we extract the \texttt{chain ID} of these rollups and request the \texttt{chain ID} using the Chainlist API.\footnote{\url{https://chainlist.org/}} The labels we gather cover 35 main rollups.

To ensure that our dataset is accurate, we compare each address with their official documents~\cite{arbitrumdocs, taiko, scroll}. To our knowledge, our rollup labels provide significantly more coverage compared to prior studies, which have only covered a few rollups~\cite{park2024impact}. Additionally, we also find that a large number of type-3 transactions are used for blobscriptions~\cite{blobscriptions}. To filter these transactions, we use the Ethscription API to collect these blobscriptions transactions.\footnote{\url{https://api.ethscriptions.com/v2/ethscriptions/}}
Utilizing this API, we retrieve a blobscription via its transaction hash. We identify and label all instances within type-3 transactions belonging to blobscriptions. Consequently, our labeling process encompasses 106{,}051 type-3 transactions used for blobscriptions. We identify all unlabeled addresses in our type-3 transaction dataset and annotate 5{,}655 Blobscription addresses following EIP-4844.

\section{Characterizing Existing Strategies}
\label{sec:builder}

To analyze how builders select transactions and how rollups verify their transactions after the upgrade (\textbf{RQ1}), we first quantify block-building strategies. We then compare the usage of calldata and blobs, to identify the transition of rollups.

\subsection{Block Building Strategies}
\label{sec:5_blocksize}
To illustrate how builders adjusted their block construction strategies after EIP-4844, we analyze the evolution of block size and its main components. \autoref{fig:avg_block_size} presents two metrics from December 2023 to August 2024: the average block size, and the size of non type-3 transactions within each block. 
As shown in~\autoref{fig:avg_block_size_block}, the overall average block size (black line) exhibits a clear upward trend. Following EIP-4844, the total block size has risen sharply from below 150 KB to nearly 400 KB, reflecting a shift in builders’ construction strategies.
Among individual builders, Flashbots and rsync (blue and green lines) display distinct adjustments: Flashbots excluded type-3 transactions in May, and then resumed the inclusion in July, suggesting its experimentation with post-upgrade strategies.

To measure the builder's varying behavior towards other non type-3 transactions after the upgrade, we calculate the size of the non type-3 transaction portion of the block(this is the space occupied by all non type-3 transactions within a block) in~\autoref{fig:avg_block_size_non-3}.
After exceeding 300 KB in January, the average block size without type-3 transactions began to decline to 80 KB by June, reducing by almost 4x. 
This confirms the efficacy of EIP-4844.
The yellow and orange lines represent the strategy of the Beaverbuild builder and Vanilla builders, who were the first to implement the strategy of decreasing the total size of other transactions within a block following EIP-4844. 
Nearly 60\% of the market share is held by these builders, and the majority of the blocks they construct are below the average size prior to May. By March, Vanilla builders transitioned to creating smaller blocks of other transactions, possibly due to not receiving a higher volume of orders and lacking the professional expertise to build blocks. Notably, starting from June, the top five market share builders (including Flashbots builder and Titan builder who previously built the largest blocks) have begun creating smaller sized blocks.

In conjunction with~\autoref{fig:avg_block_size_block}, this finding suggests that the builders are making increasing room for type-3 transactions by reducing the space for others.
After June, the size of this portion stabilized below 80 KB, indicating that the growing prevalence of type-3 transactions has effectively constrained the space available for type-2 transactions.

\subsection{Layer-2 Data Verification}
\label{sec:rollup}

\begin{figure}[!bpt]
    \centering
    \includegraphics[width=1\linewidth]{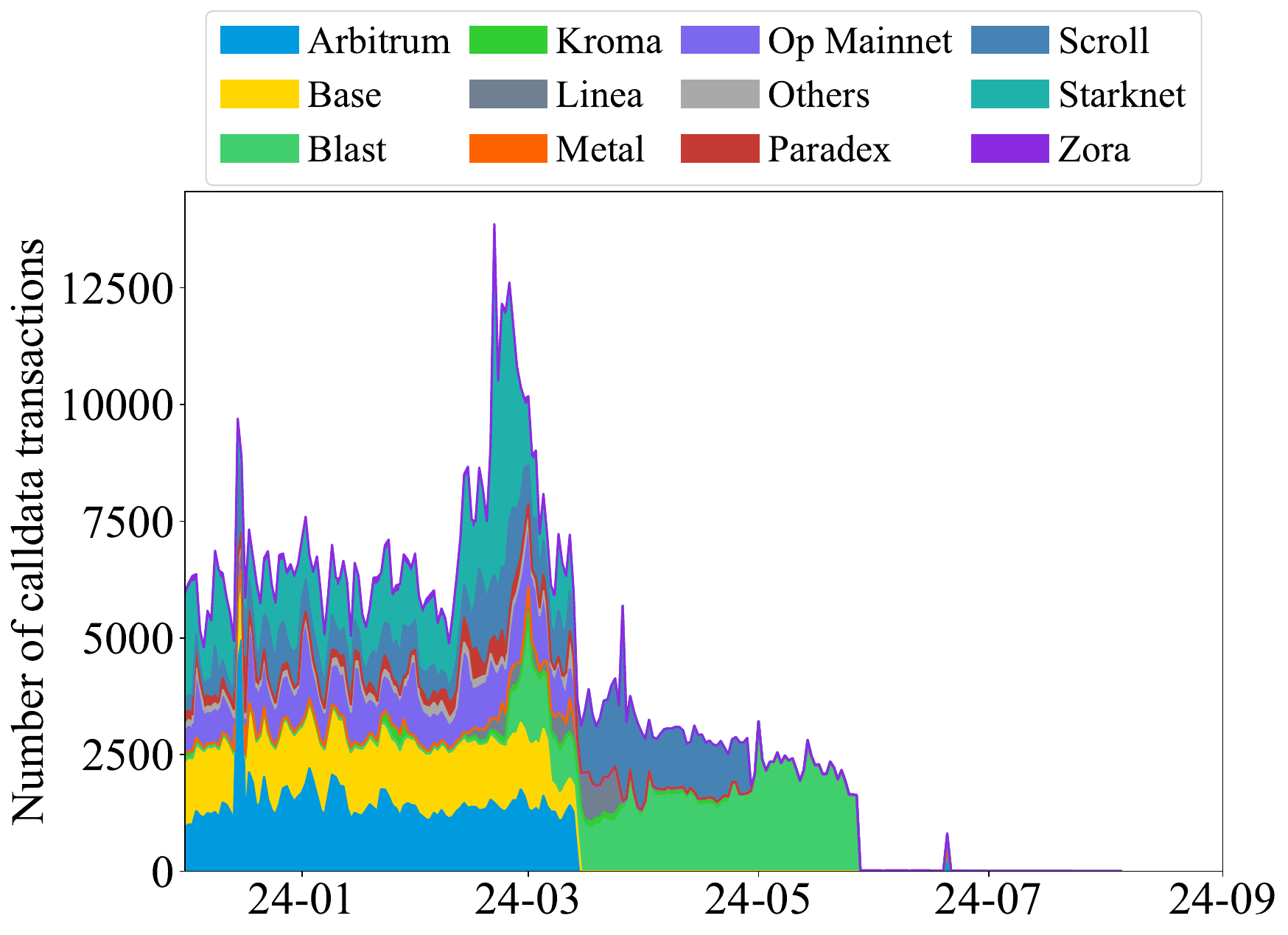}
    \Description{.}
    \caption{The landscape of calldata usages.}
    \label{fig:calldata_landscape}
\end{figure}

\begin{figure}[!bpt]
    \centering
    \includegraphics[width=1\linewidth]{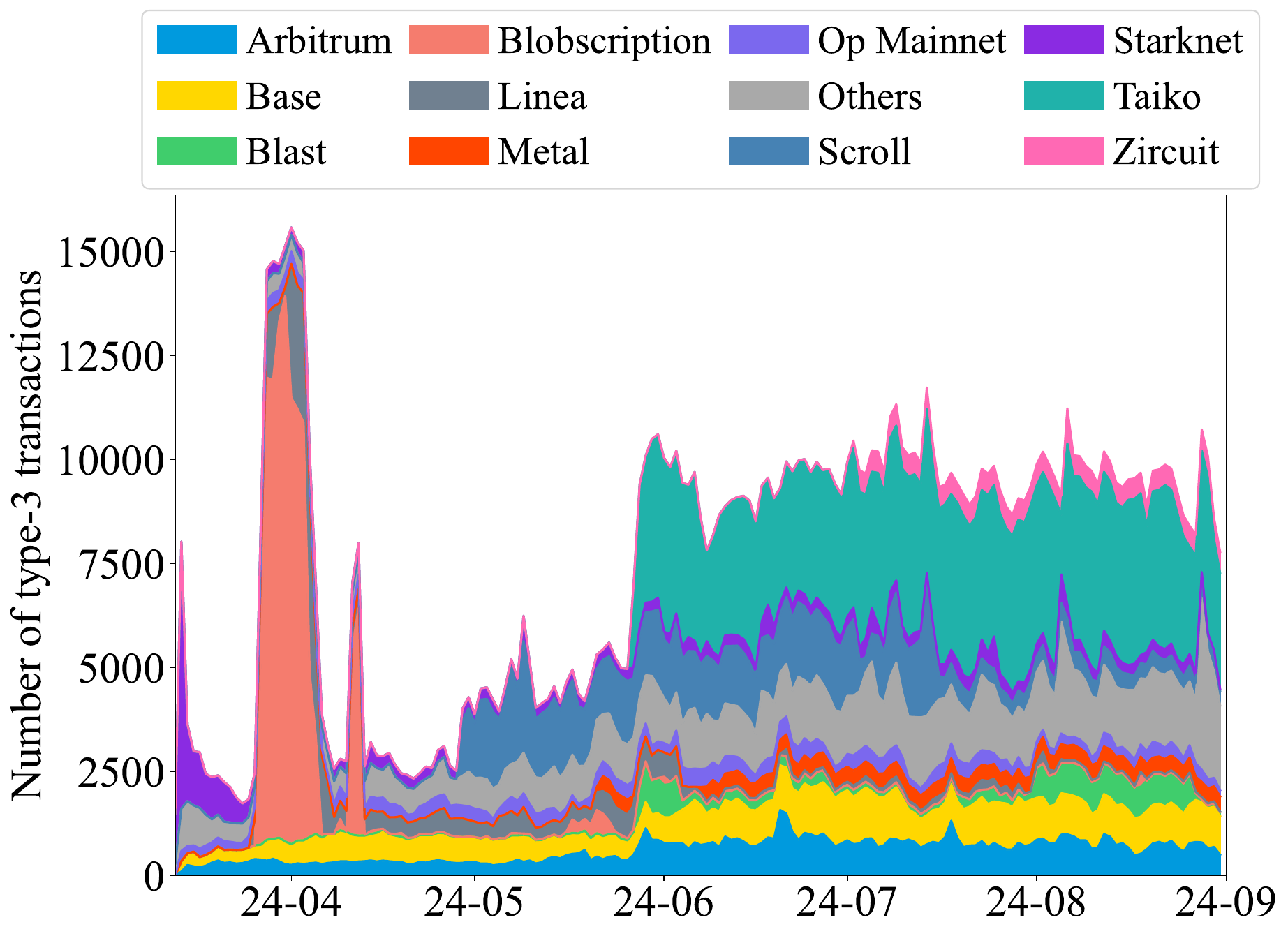}
    \Description{.}
    \caption{The landscape of blob usages.}
    \label{fig:rollup_landscape}
\end{figure}

While builders modify their block building strategies in response to EIP-4844, rollups have simultaneously adjusted their data submission strategies. \autoref{fig:calldata_landscape} plots the daily use of calldata pre- and post-EIP-4844. Before the upgrade, rollups use calldata to finish their verification. Owing to blobs' lower transaction fees, rollups have increasingly adopted blobs to validate layer-2 transactions. The use of calldata gradually declined from over 7{,}500 transactions per day until it nearly disappeared. After the upgrade, in total 1{,}336{,}822 type-3 transactions have been sent to the Ethereum mainnet. \autoref{fig:rollup_landscape} visualizes the daily share of type-3 transactions by the dominant rollups and another specific purpose (Blobscription) after EIP-4844, in terms of the total number of blob-carrying transactions. 
The amount of daily transactions carrying blobs has increased to approximately 10{,}000 in June. We notice that the number of transactions for each rollup varies over time, with some rollups showing significant spikes or drops. One of the most noticeable spikes is in Blobscriptions, which began hitting the market at the end of March, reaching its peak before decreasing dramatically. The purpose of Blobscriptions is not to validate rollup data, but rather to publish it. Other rollups show varying increases and decreases in usage, which is related to the choices made by each rollup. 
The use of type-3 transactions has been increasing after EIP-4844, with more rollup applications using blobs to validate data. Instead of immediately using a blob to transfer data, some rollups (e.g., Blast and Scroll) only started adopting type-3 transactions in May of this year. However, there are still many rollups (such as Arbitrum and Base) that have been using type-3 transactions since EIP-4844.

In conclusion, existing rollups have largely transitioned from using calldata to blob in type-3 transactions, as EIP-4844 significantly reduced transaction fees. Daily calldata usage has decreased from approximately 7{,}500 to nearly none, while blob data has simultaneously surged from zero to around 10{,}000 per day. 
This upgrade has arguably created a new game between block builders and rollups, yet the dynamics and strategies of this game remain poorly understood.

\section{Understanding Optimal Strategies}
\label{sec:Game Model}

The previous section illustrates entities' strategic transitions post EIP-4844. We next model the game between builders and rollups to determine the equilibrium priority fee chosen by builders, and the optimal blob-carrying strategy adopted by rollups (\textbf{RQ2}). We start with a single-round model to describe the behavior of the builder, and then extend this model to multiple rounds.~\autoref{table:notation} summarizes the commonly used notations used throughout the paper.

\begin{table}[!bpt]
    \centering
    \caption{The frequently used notations.}
    \begin{tabular}{@{}>{\centering\arraybackslash}p{1.3cm} p{6.9cm}@{}}
        \toprule
        \textbf{Notation} & 
        \textbf{Description}  \\
        \midrule
        $m$ & The type-2 transaction in public mempool. \\
        $n$ & The type-3 transaction.\\
        $q$ & Quantity of type-2 transactions in one round.\\
        $k$ & Quantity of type-3 transactions in one round.\\
        $\mathbb{M}$ & A set of type-2 transaction $\mathbb{M} = \{m_1,m_2,\ldots,m_q\}$. \\
        $\mathbb{N}$ & A set of type-3 transaction $\mathbb{N} = \{n_1,n_2,\ldots,n_k\}$. \\
        $s^{(2)},p^{(2)}$ &  Size and priority fee for the type-2 transaction. \\
        $s^{(3)},p^{(3)}$ &  Size and priority fee for the type-3 transaction. \\
        $r$ &  Floor price for type-3 transactions.\\
        $V \& \tilde{V}$ & Revenue of type-3 transaction and the initial value. \\
        $\rho(t)$ & Time-dependent discount factor.\\
        $\pi(V)$ & Winning probability of the type-3 transaction.\\
        $u$ & Utility of the type-3 transaction.\\
        \bottomrule
    \end{tabular}
    \label{table:notation}
\end{table}

\subsection{Single-Round Model}
\label{sec:efficient_strategy}
For builders, the single-round model characterizes behavior, as builders are myopic and focus only on maximizing utility within each round. Transactions in the mempool are selected to fill block space, where profits are obtained through the `maxPriorityFeePerGas' field (for convenience, we refer to this as the `priority fee'). 
For rollups, type-3 transactions are submitted without observing others’ priority fee, resembling a sealed-bid setting~\cite{vickrey1961counterspeculation}. The selection process is formalized as a first-price sealed-bid auction, in which rollups compete for block space and pay priority fees as bids. The decision between type-2 and type-3 transactions also becomes a critical consideration, while builders determine which transactions to include into block.

Let $\mathbb{M} = \{m_1, m_2, \ldots, m_q\}$ denote $q$ type-2 transactions remaining in the public mempool, and let $\mathbb{N} = \{n_1, n_2, \ldots, n_k\}$ denote $k$ type-3 transactions. For builders, research indicates that price and block size are the primary factors of concern~\cite{wu2024strategic, wahrstatter2023time}. For each type-2 transaction $m_i \in \mathbb{M}$, its size is denoted as $s_i^{(2)}$ and its priority fee as $p_i^{(2)}$. For type-3 transaction $n_j \in \mathbb{N}$, the size is denoted as $s_j^{(3)}$ and the priority fee as $p_j^{(3)}$.

The floor priority fee $r$ of type-3 transactions represents the maximum fee attainable by selecting type-2 transactions of the same size. Formally, the floor price $r_{j}$ is defined as

\begin{equation}
\begin{aligned}
    r_{j} = \max \sum_{m_i \in \mathbb{M}} p_i^{(2)},\\
    \text{s.t.} \sum_{m_i \in \mathbb{M}}  s_i^{(2)} \leq s_j^{(3)}.
\end{aligned}
\end{equation}

For rollups, the revenue of $n_j$ is denoted as $V_{j}$. For tractability, $V_{j}$ is assumed to be independently drawn from a distribution with cumulative density function $F(\cdot)$ in auction games~\cite{krishna2009auction,qin2022quantifying}.
The quantity $k$ of type-3 transactions and the distribution $F(\cdot)$ are assumed to be prior knowledge of both builders and rollups. It is further assumed that sufficient block space is available, such that block size constraints do not materially influence the selection probability, with empirical evidence provided in~\autoref{app:blob}. The selection probability is denoted as $\pi(V_{j})$. Consequently, given the floor price $r_{j}$, the expected utility of the type-3 transaction $u_{j}$ is

\begin{equation}
\mathbb{E}\left[u_{j}\right]=\begin{cases}\pi(V_{j})\left(V_{j}-p_{j}^{(3)}\right), \quad &p_{j}^{(3)}\geq r_j. \\
0, \quad &p_{j}^{(3)}<r_j.
\end{cases}
\end{equation}

At symmetric equilibrium, each type-3 transaction selects $p_{j}^{(3)}$ to maximize $u_{j}$. For tractability, $V_{j}$ is always assumed to be independently drawn from a uniform distribution as the prior knowledge of rollups in auction games, i.e., $V_{j} \sim U(0,V_{max})$~\cite{krishna2009auction,qin2022quantifying}. 
Under this simplifying assumption, the priority fee for type-3 transactions accepted by builders satisfies:
\begin{equation}
\label{selectionstrategy}
p^*=\max\left(r_{j},\frac{k-1}{k}\max_j V_{j}\right), 1 \leq j \leq k.
\end{equation}

In conclusion, we derive the transaction selection strategy of builders when the game reaches the Bayesian Nash equilibrium, and the proof for Eq.~\autoref{selectionstrategy} can be found in~\cite{easley2010networks}. It can be observed that, as the number of type-3 transactions increases, the priority fee increases and gradually approaches the revenue that the rollups derive from these transactions. The optimal strategy for builders is therefore comparing the highest priority fee of type-3 transactions with its floor price and then select the higher one.

\subsection{Multiple-Round Model}
\label{sec:multiple model}

After the builder's strategy and equilibrium situation have been analyzed, the discussion shifts to rollups. When multiple blobs must be submitted within one round, rollups need to determine how many blobs to bundle into each type-3 transaction. This consideration is directly related to how rollups minimize the cost per blob.

For rollups, if a type-3 transaction was not selected, it will remain in the mempool and continue to participate in subsequent builder-rollup auctions. Rollups therefore retain the opportunity for their transactions to be included in later rounds. However, type-3 transactions are time sensitive. Fraud proofs in optimistic rollups must be submitted within a strict challenge window, while validity proofs in ZK-rollups directly determine the finalization of batched Layer-2 transactions on the mainnet~\cite{scroll, arbitrumdocs, taiko}. To capture this sensitivity, a time-dependent discount factor $\rho(t)$ is introduced, where $t$ denotes discrete rounds. The discount reflects the decline of revenue as type-3 transactions wait for inclusion in later rounds. The single-round first-price sealed-bid auction model which we have talked about before is now extended into a multi-round game, and the revenue of type-3 transaction $n_j$ is expressed as
\begin{equation}
    V_{j} = \rho(t) \tilde{V}_{j},
\end{equation}
where $\tilde{V}_{j}$ denotes the revenue in the initial round $t=0$. The function $\rho(t)$ is strictly decreasing, with $\rho(0)=1$ and $\rho^{\prime}(t)<0$. Therefore, the expected utility of multiple rounds becomes a geometric distribution form as
\begin{equation}
    \mathbb{E}[u_j]=g(\tilde{V_{j}})=\frac{\pi(\tilde{V}_{j})}{k}\tilde{V_{j}}\sum_{t=1}^\infty(1-\pi(\tilde{V}_{j}))^t\rho(t).
\end{equation}

The analysis starts with a rollup submitting two blobs, serving as the foundational scenario for further cases. In the first strategy, two blobs are bundled into a single type-3 transaction $n_a$. In the second strategy, the two blobs are submitted separately as type-3 transactions $n_b$ and $n_c$, with $V_{a} = V_{b} + V_{c}$. The expected utility of bundling is $\mathbb{E}[u_{n_a}] = g(V_{b} + V_{c})$, while the expected utility of splitting becomes $\mathbb{E}[u_{b}] + \mathbb{E}[u_{c}] = g(V_{b}) + g(V_{c})$. The comparison thus depends on the curvature of $g(V)$. 

\begin{algorithm}[!bpt]
\setlength{\hsize}{0.95\linewidth}
\small
\SetAlFnt{\small}
\SetAlCapFnt{\small}
\SetAlCapNameFnt{\small}
\SetAlgoLined
\SetKw{KwForIn}{in}
\SetKwProg{Fn}{Function}{:}{end}
\SetKwProg{Alg}{Algorithm}{:}{end}
\SetKwFunction{FMain}{max\_tips\_greedy}
\SetKwFunction{Sort}{Sort}
\KwIn{The type-2 transactions remaining in the public mempool $\mathbb{M}$ , the size \& price of type-2 transactions $s^{(2)} \& p^{(2)}$, the size of type-3 transactions $s^{(3)}$;}

\For{\text{$m_i$} in $\mathbb{M}$}{
    $\text{ratio} \gets \text{$p^{(2)}_i$} / \text{$s^{(2)}_i$}$\;
}
$\mathbb{M} \gets \Sort\left(\mathbb{M}, \text{by ratio in descending order}\right)$\;

$\text{total\_size} \gets 0$\;
$\text{total\_tips} \gets 0$\;
$\text{selected\_list} \gets []$\;
\For{\text{$m_i$} in $\mathbb{M}$}{
    \If{$\mathrm{total\_size + \text{$s^{(2)}_i$}} \leq s^{(3)}$}{
        $\mathrm{total\_size} \gets \mathrm{total\_size} + \text{$s^{(2)}_i$}$\;
        $\mathrm{total\_tips} \gets \mathrm{total\_tips} + \text{$p^{(2)}_i$}$\;
        $\mathrm{selected\_list}.\text{append}(\text{$m_i$})$\;
    }
}
$\mathbb{M} \gets \mathbb{M}.\text{Drop} (\mathrm{selected\_list})$\;
\KwRet{$\text{total\_tips}, \text{selected\_list}$}\;

\caption{Greedy Algorithm}
\label{alg:greedy}
\end{algorithm}


From the second derivative of $g(V)$, it is observed that under conditions of sufficient block space, large $k$, and a slow decrease in $\rho(t)$, $\pi(V)$ increases rapidly, resulting in a convex $g(V)$ with $\pi^{\prime\prime}(V)$ remaining positive under a uniform distribution. Empirical evidence in~\autoref{app:blob} indicates that type-3 transactions carrying multiple blobs do not experience significant additional latency, which aligns with these conditions. Therefore, bundling multiple blobs into a single type-3 transaction is generally the more advantageous strategy for rollups.

In conclusion, bundling several blobs into one type-3 transaction is expected to deliver higher utility in practice. Starting from the two-blob case, the model extends naturally to the general case of multiple blobs, and the conclusion remains consistent. 
\section{Existing Market Efficiency}
\label{sec:existing_inefficiency}
We derive the equilibrium condition and optimal strategy for builders and rollups in~\autoref{sec:Game Model}. In this section, we pay attention to whether the current market employs optimal strategies (\textbf{RQ3}). If current strategies do not align with the model conclusions, we consider them as economically inefficient.

\subsection{Builder Efficiency}
\label{sec:efficiency}
\begin{figure*}[!bpt]
    \centering
    \begin{subfigure}{0.49\textwidth}
        \centering
        \includegraphics[width=\linewidth]{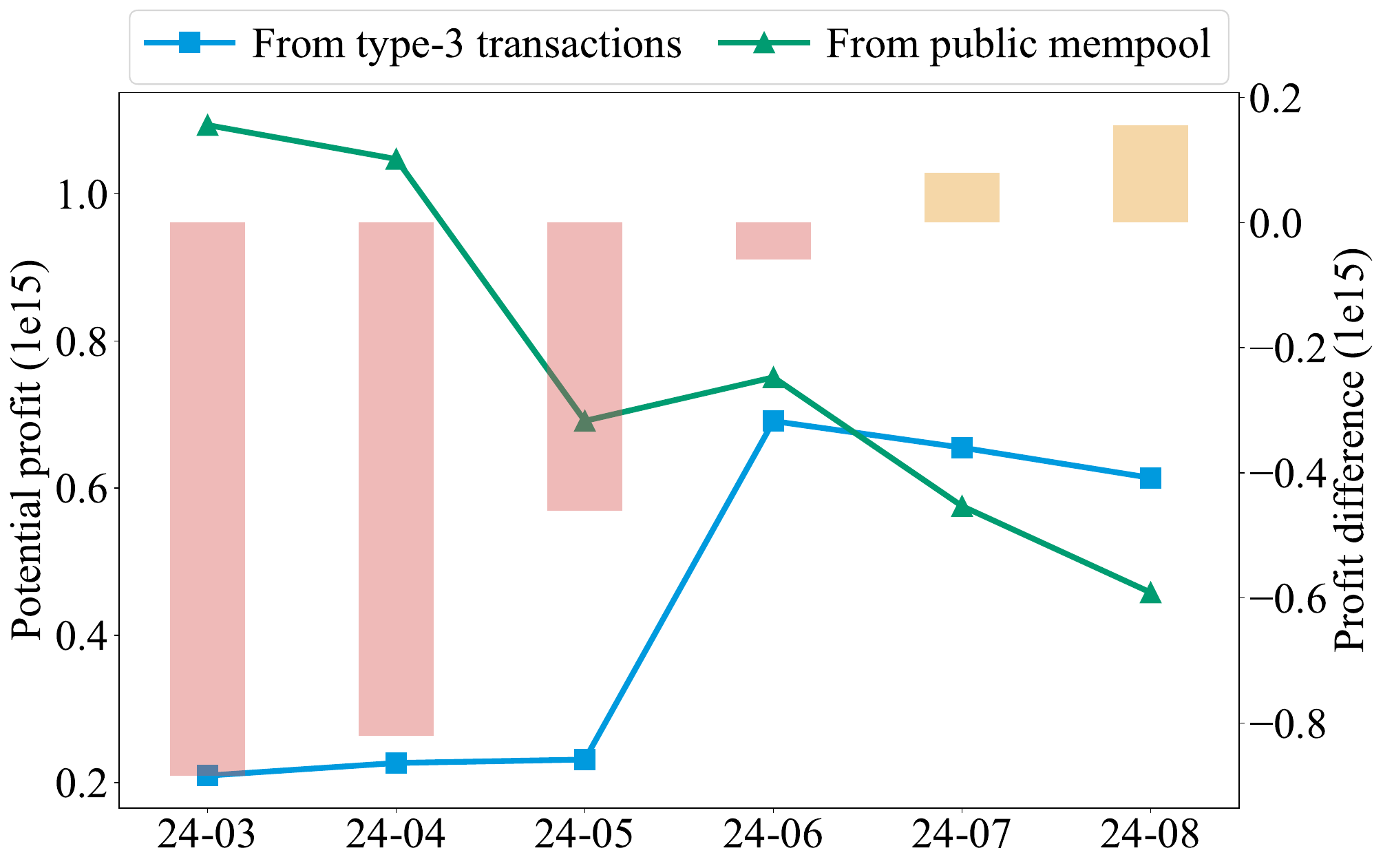}
        \caption{The proportion of inefficient blocks per builder.}
        \Description{.}
        \label{fig:efficient_trend_builder}
    \end{subfigure}
    \hfill
    \begin{subfigure}{0.5\textwidth}
        \centering
        \raisebox{-0.6cm}{%
            \includegraphics[width=\linewidth]{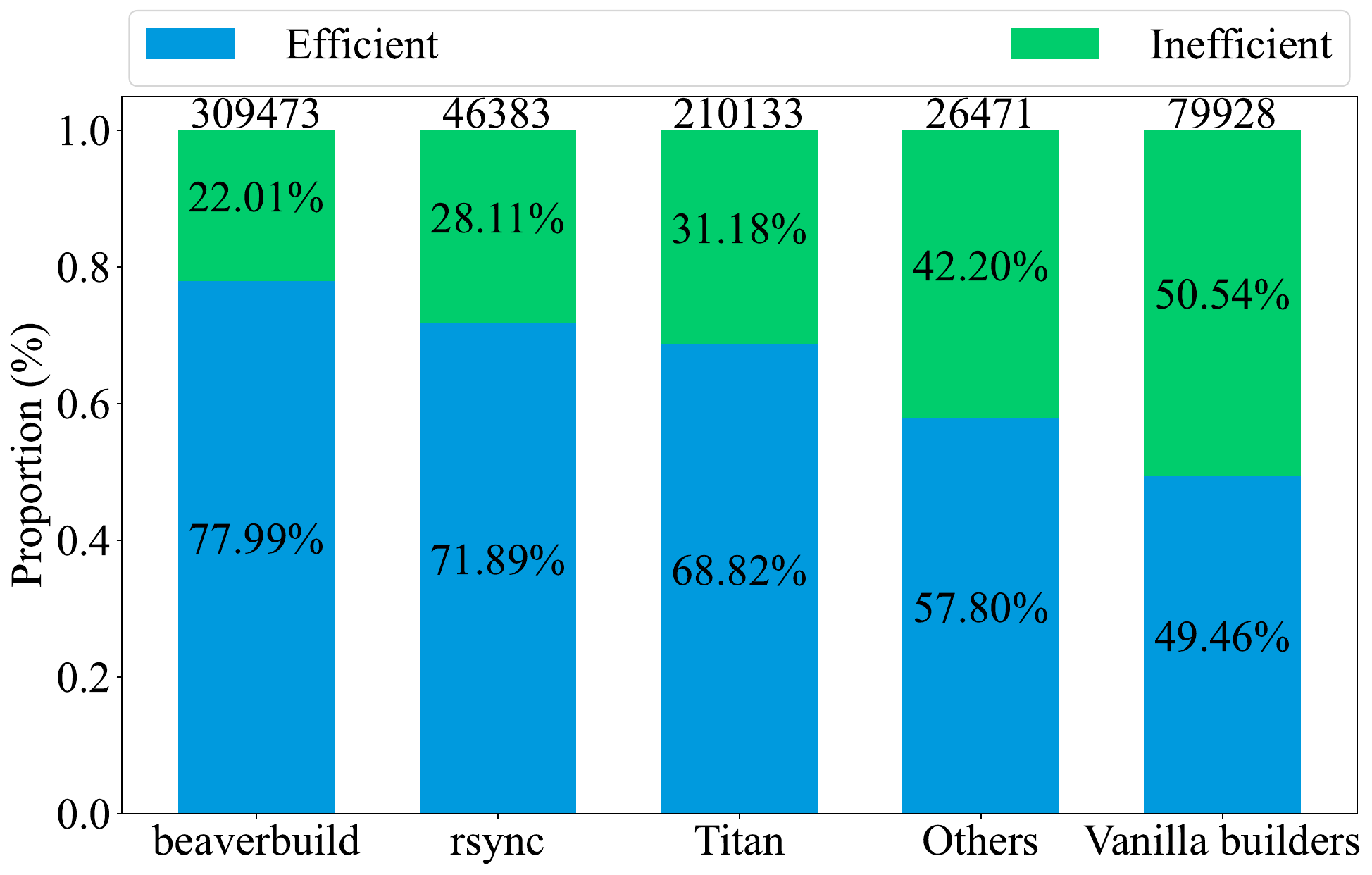}%
        }
        \caption{The profit of type-3 transactions and public mempool over time.}
        \Description{.}
        \label{fig:efficient_proportion_bar}
    \end{subfigure}
    \Description{.}
    \caption{The efficiency of builders. The left figure shows the profit gap and the right figure illustrates the different efficiencies among distinct builders.}
    \label{fig:builder_efficiency}
\end{figure*}

\begin{table}[!bpt]
    \centering
    \caption{The overall inefficient block distribution over time.}
    \begin{tabular}{l c c c c}
        \toprule
        \textbf{} & \textbf{Inefficient} & \textbf{Total} & \textbf{Inefficient Proportion} \\
        \midrule
        \textbf{March} & 26,450 & 47,604 & 55.56\% \\
        \textbf{April} & 41,779 & 84,631 & 49.37\% \\
        \textbf{May} & 47,351 & 105,382 & 44.93\% \\
        \textbf{June} & 31,540 & 139,529 & 22.60\% \\
        \textbf{July} & 27,965 & 145,726 & 19.19\% \\
        \textbf{August} & 23,139 & 149,515 & 15.48\% \\
        \textbf{Overall} & \textbf{198,224} & \textbf{672,388} & \textbf{29.48\%} \\
        \bottomrule
    \end{tabular}
    \label{table:month_efficient}
\end{table}
To measure whether builders are adopting the transaction selection strategy, we employ a greedy algorithm considering size and profit to measure whether the priority fee given to each type-3 transaction is higher than the floor price defined in~\autoref{sec:efficient_strategy} during the concurrent round. To filter out transactions that builders have simulated as likely to fail, we filter out transactions in the public mempool with a delay of more than 100 seconds to make the calculation more accurate. We provide the pseudocode of the greedy algorithm in~\autoref{alg:greedy}, considering the size and priority fee per transaction, calculating the priority fee ratio for each transaction and sorting by the ratio before processing.

\autoref{fig:efficient_trend_builder} plots the priority fee that builders extract from type-3 transactions versus the priority fee that can be extracted from type-2 transactions in the same round (from March to August 2024). We compute the priority fee difference, and indicate this value with red and yellow bars. A red bar indicates that this difference is negative, which suggests that the builders could have made more profit from the public mempool, but they chose to include the less profitable type-3 transactions in the block. Conversely, if the bar is yellow, it indicates that the value is positive and the builder made more priority fee from type-3 transactions. It is clear that after EIP-4844, the builder misses significant priority fees from the public mempool. As~\autoref{fig:efficient_trend_builder} shows, in March, builders could have potentially made more than 5x more profit from the public mempool as they could from type-3 transactions. However, this loss diminishes over time and reverses in July, when the builder makes more profit from type-3 transactions than it is able to make from the public mempool. 

We refer to the block that contains transactions that have not reached the maximum priority fee as an \emph{inefficient block}. \autoref{table:month_efficient} illustrates the details from March to August, including the number of efficient and inefficient blocks and their proportion out of the 636{,}981 blocks containing type-3 transactions. 
We find that 196,090 blocks are inefficient, covering 29.48\% of all blocks. Encouragingly, the percentage of inefficient blocks dropped from 55.56\% immediately after EIP-4844 to 15.48\% by August, indicating a trend toward a more efficient block building market. However, 15.48\% remains a non-negligible share, underscoring that there is still considerable room for improvement. 

\autoref{fig:efficient_proportion_bar} shows the ratio of efficient blocks among major builders during the same period. The y-axis shows the proportion of efficient and inefficient blocks. Beaverbuild has a market share of almost 50\%, and is the most efficient builder compared to other builders. In contrast, non-specialist builders, Vanilla builders who do not specialize in block construction, have nearly 30\% more inefficient blocks than beaverbuild, and more than half of the blocks they construct are inefficient.
\begin{figure}[!bpt]
    \centering
    \includegraphics[width=1\linewidth]{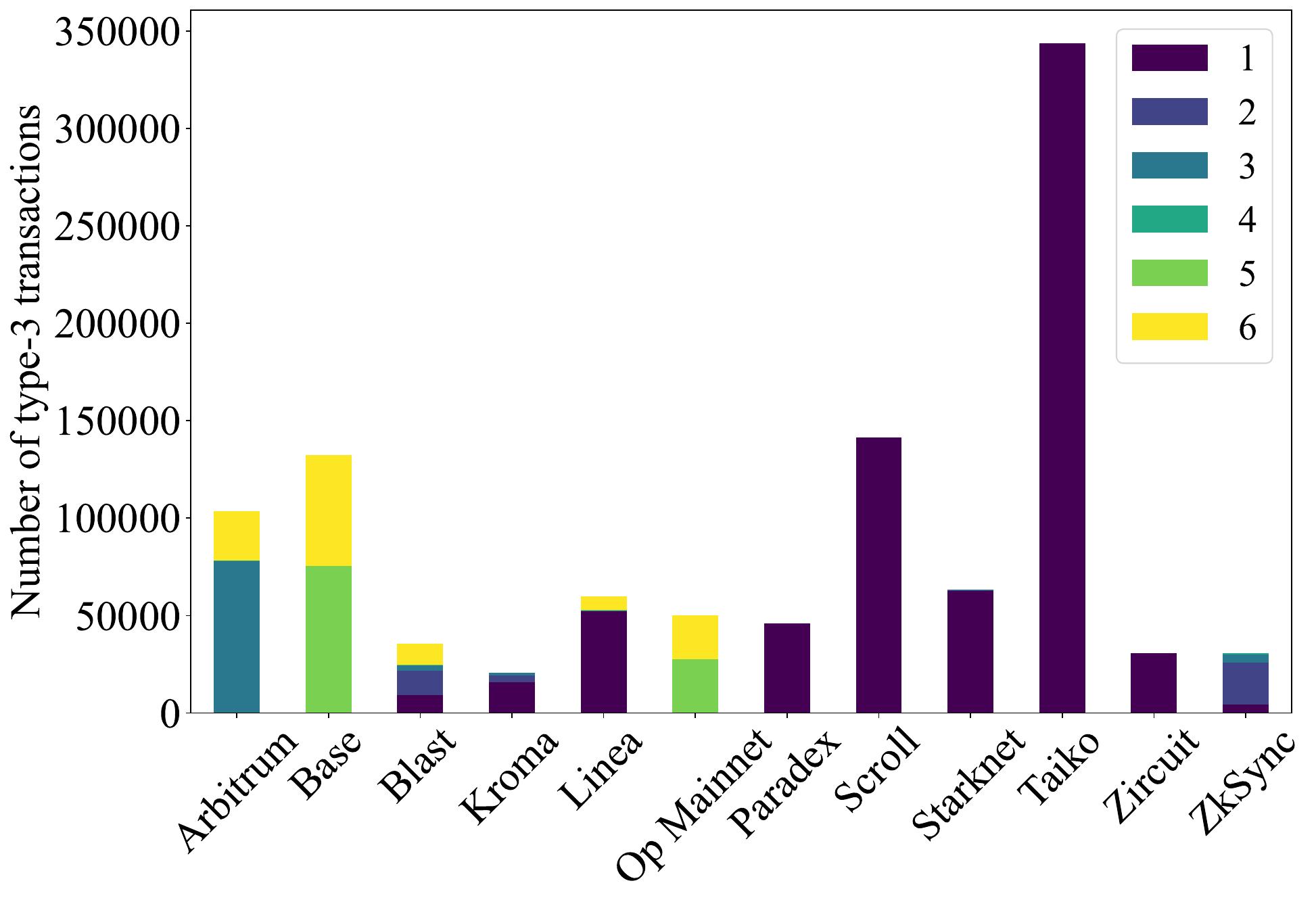}
    \Description{.}
    \caption{The blob packing strategies among rollups.}
    \label{fig:blob_bar}
\end{figure}

\subsection{Rollup Efficiency}
\begin{figure}[!bpt]
    \centering
    \includegraphics[width=0.95\linewidth]{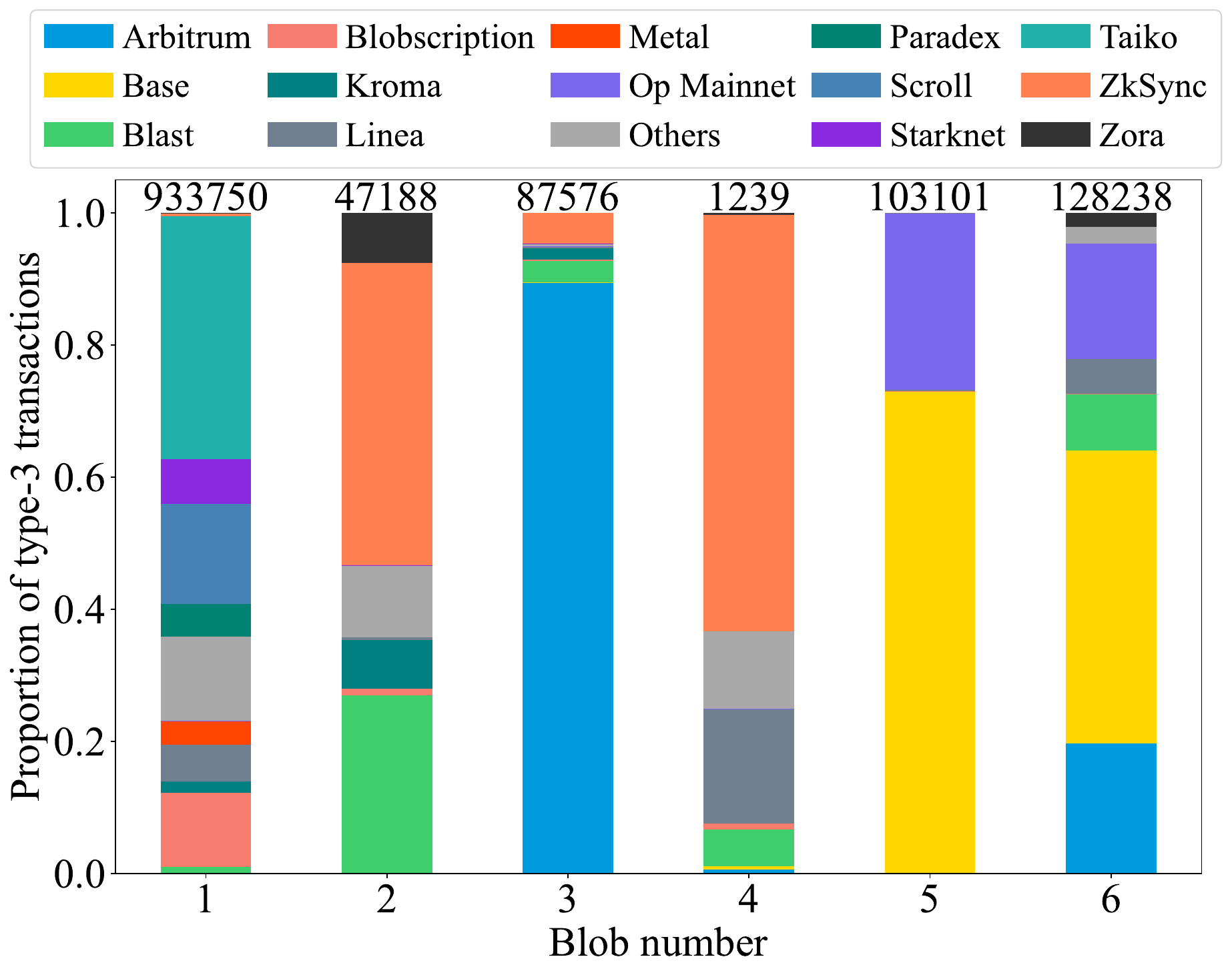}
    \Description{.}
    \caption{The landscape of blob usage strategies.}
    \label{fig:t3_landscape}
\end{figure}
\autoref{fig:blob_bar} plots the blob packing strategy for each mainstream rollup from the introduction of blob to 31 August 2024. The different colors represent the number of blobs carried in type-3 transactions and the Y-axis represents the total number of transactions. It is clear that rollups use different strategies in submitting blobs. Of the 343,468 transactions issued by Taiko, it puts only one blob into the type-3 transaction, and the same strategy is used by Scroll, Starknet, and Zircuit. In contrast, Base and Op Mainnet, which have been using blobs to send data, typically fill each transaction to the maximum capacity --- up to six blobs per transaction, which is the current limit. In addition, there are some rollups that do not limit how many blobs they carry in a single transaction (such as Linea and Blast). From our derivation in~\autoref{sec:multiple model}, the bundling strategy adopted by rollups such as Arbitrum and Base is more advantageous than alternative approaches.

We also show, in~\autoref{fig:t3_landscape}, the percentage of type-3 transactions issued by different rollups that carry a specific number of blobs. 
Most rollups use relatively fixed strategies, as signified by the existence of a color in a single bar. 69.85\% of the transactions in the network carry only one blob. Despite this, it is \emph{not} the optimal strategy. Additionally, transactions carrying 2 to 4 blobs are the least common. 
In conclusion, while the optimal strategy for rollups is to bundle multiple blobs into one type-3 transaction, it is only adopted by Arbitrum and Base. The optimal strategy is not yet widely used in the current market. 
\section{Related Work}
\label{sec:related}
\parhead{Ethereum Upgrades.~}
Prior studies have explored several upgrades related to Ethereum, including the proposal of ERC-20 tokens~\cite{lin2024denseflow, wu2023know, zhao2021temporal, wu2023defiranger, kong2023defitainter}, NFT~\cite{huang2024unveiling, ito2024investigations, xiao2024centralized}, the London hard fork~\cite{liu2022empirical, leonardos2023optimality}, and the transition from PoW to PoS~\cite{heimbach2023defi, silva2020impact, chi2024remeasuring, mclaughlin2023large}. After these updates, detection frameworks have been proposed to address the security issues posed by cryptocurrency tokens~\cite{hu2024piecing, victor2021detecting, li2022ttagn}. Hu et al.~\cite{hu2023bert4eth} developed a model to detect Ethereum fraud, while Zhou et al.~\cite{zhou2024artemis} examined the use of NFT for arbitrage and airdrop acquisition and revealed that it is driven by a small number of dominant players. 
Regarding the Ethereum merge event, Heimbach et al.~\cite{heimbach2023defi} quantified and analyzed hard-fork arbitrage in lending protocols. Our work complements these studies with a systematic analysis of EIP-4844 and provides valuable insights into the emerging blob market.

\parhead{Builder Behaviors.~}
Several previous works, such as~\cite{oz2023time, ovezik2024sok, gupta2023centralizing, wang2025private}, investigated builder behaviors in PBS. The landscape exhibited that builders had a significant trend of centralization~\cite{heimbach2023ethereum}. Likewise, Wahrstatter et al.~\cite{wahrstatter2023time} showed how vertical integration could potentially extract more value and further centralize block construction. \"Oz et al.~\cite{oz2024wins} explored the relationship between the profit and market share, and how private order flow intensified the centralization of the builder market. In addition, Yang et al.~\cite{yang2024decentralization} focused on competition and efficiency in MEV-Boost auctions of builders, and Heimbach et al.~\cite{heimbach2024non} quantified the non-atomic arbitrage utilized by builders in the block-building process. Aside from this, blockchain censorship was defined to measure the censorship of builders~\cite{wahrstatter2024blockchain, wang2023first}, demonstrating security pitfalls. However, no prior work has explored builder strategies from the perspective of block sizes and selection choices. To the best of our knowledge, we are the first to measure builders' strategies towards distinct transactions and evaluate their efficiency, offering crucial insights for builders to refine their block construction.

\parhead{Rollup Analysis.~}
Prior studies on the performance of the rollup ecosystem have explored its dynamics and potential future advances~\cite{palakkal2024sok, chaliasos2024analyzing}. 
Torres et al.~\cite{torres2024rolling} analyzed the process of MEV extraction within Layer-2 rollups, while Avil et al.~\cite{khalil2024parole} investigated the arbitrage of NFT that occurs in rollups. 
Gogol et al.~\cite{gogol2024writing} examined the effects of inscriptions on rollups. They found that the rise in meme transactions leads to reduced gas fees for rollups. Stefanos et al.~\cite{chaliasos2024analyzing} provided insight into zero-knowledge rollups and developed a structured approach for their evaluation. Closest to our work is the study by Crapis et al.~\cite{crapis2023eip}, where they introduced an economic framework to examine EIP-4844 and its impact on rollups for reducing the cost of mainnet data. 
In comparison, our research provides the first rollup behavior evaluation and devises the optimal strategy, which is useful for rollups to optimize their data verification strategies in the post-EIP-4844 era.
\section{Conclusion}
\label{sec:conclusion}
In this work, we have presented the first study of the interactions between builders and rollups introduced by Ethereum’s Dencun upgrade (EIP-4844). By combining a large-scale dataset of over 319 million transactions, with a formal game-theoretic model, we shed light on the new economic dynamics between block builders and rollup operators in the blob marketplace. Our analysis reveals that both builders and rollups often employ suboptimal strategies, leading to significant inefficiencies and measurable losses. 
It is worth noting that over 29 \% of the blob-containing blocks are built suboptimally. 
These findings highlight the evolving strategic complexity of post-upgrade block construction and blob submiting markets. Future research may extend this work by exploring long-term equilibrium behaviors, the impact of emerging protocol upgrades, and the broader implications for market design.

\newpage
\bibliographystyle{ACM-Reference-Format}
\bibliography{ref}

\newpage
\appendix
\section{Blob Quantity and Delay}
\label{app:blob}
\begin{figure}[!bpt]
    \centering
    \includegraphics[width=1\linewidth]{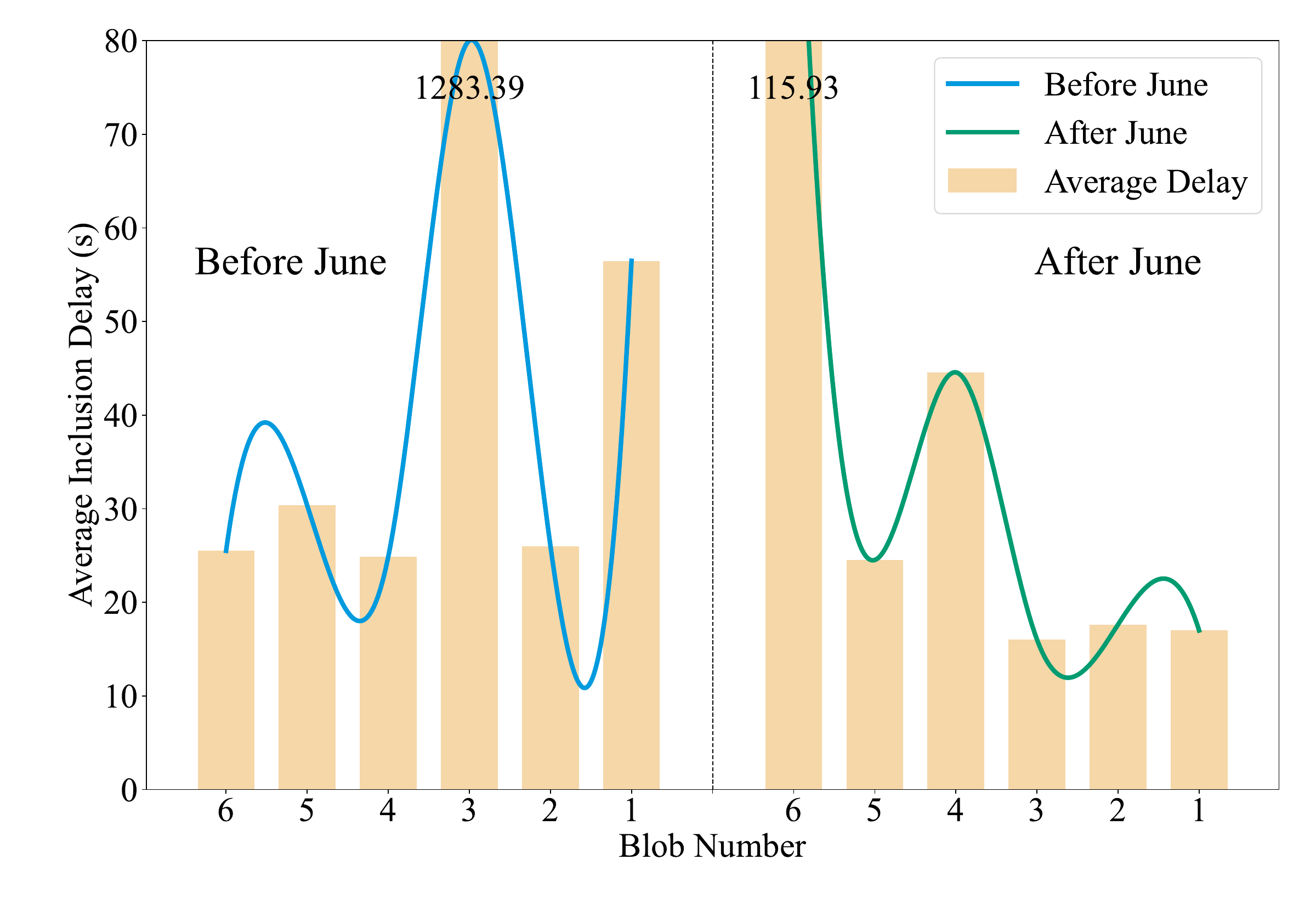}
    \Description{Inclusion delay and blob number.}
    \caption{Inclusion delay and blob number.}
    \label{fig:bar_delay_blob_number_merge_fit}
\end{figure}

Unlike ordinary auction games, the selection probability of type-3 transactions may also depend on transaction size. Hence, before defining the winning probability, it is necessary to evaluate whether block size constraints affect the inclusion probability of type-3 transactions. \autoref{fig:bar_delay_blob_number_merge_fit} presents the relationship between the number of blobs (transaction size) and the average inclusion latency of type-3 transactions after EIP-4844, with a fitted trend for the observed delays. The y-axis represents the inclusion delay, and the x-axis indicates the number of blobs per transaction. Bar colors, ranging from blue to green, correspond to transaction counts. Empirical evidence shows that, after June, type-3 transactions containing between 1 and 5 blobs incur negligible additional latency.

\section{Inefficient Strategies of Rollups}

\begin{figure}[!bpt]
    \centering
    \includegraphics[width=1\linewidth]{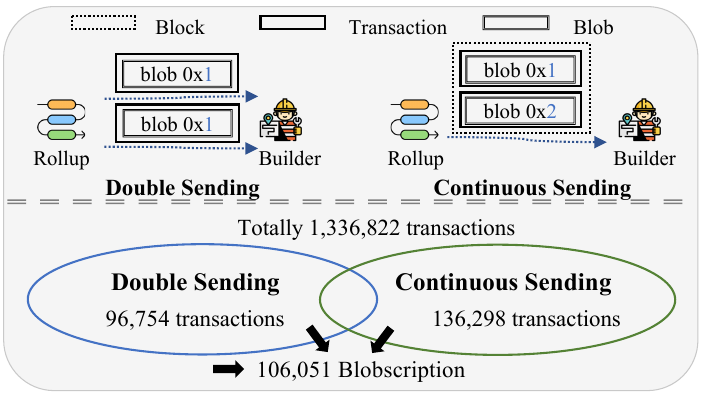}
    \Description{.}
    \caption{The behaviors and amount of inefficient strategies.}
    \label{fig:non_efficient}
\end{figure}

\begin{figure}[!bpt]
    \centering
    \includegraphics[width=1\linewidth]{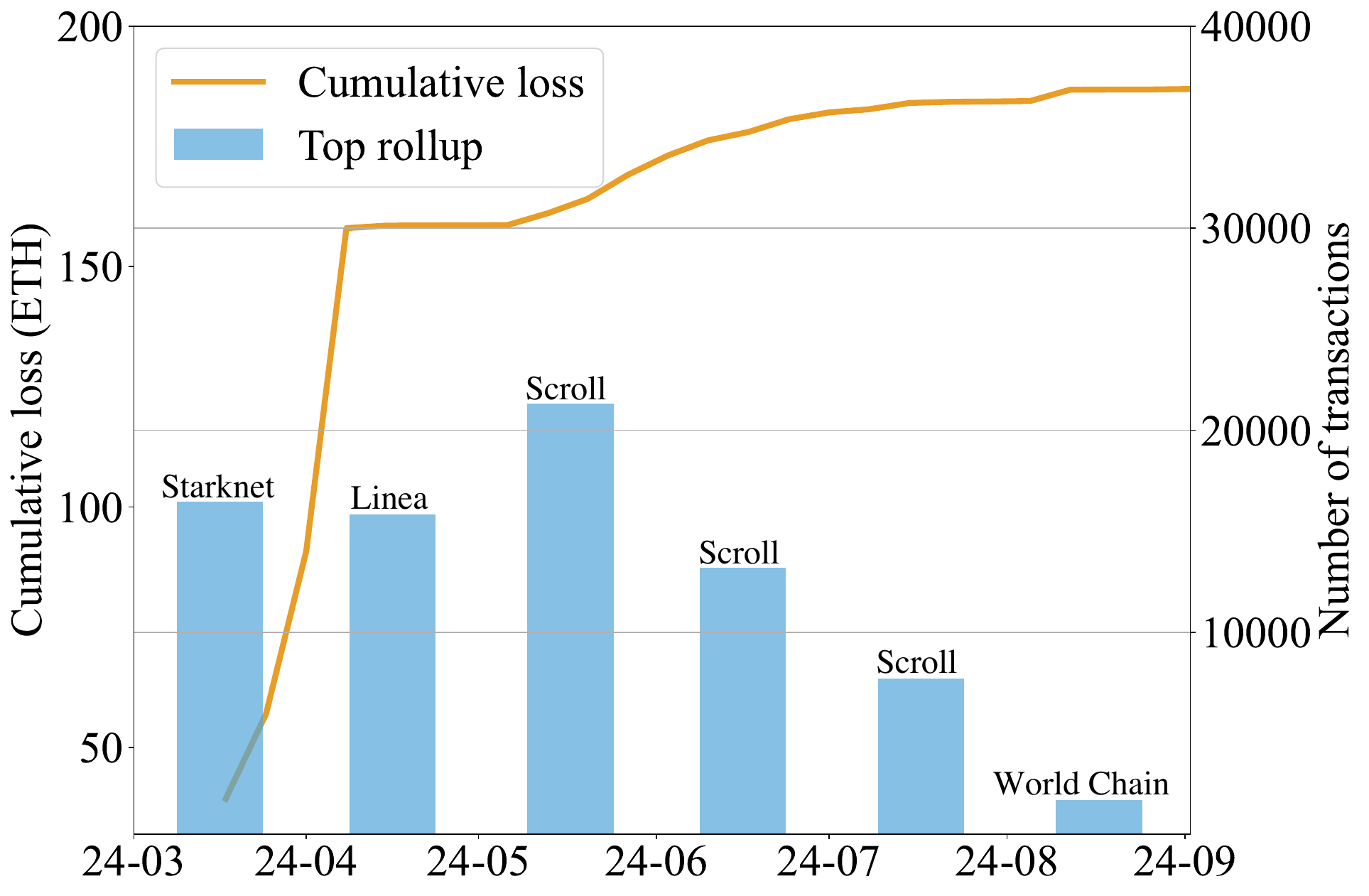}
    \Description{.}
    \caption{The cumulative loss of inefficient strategies..}
    \label{fig:cs_loss}
\end{figure}

To analyze whether rollups are using type-3 transactions and blobs efficiently, we analyze their blob-carrying strategies and identify two existing inefficient strategies, Continuous Sending and Double Sending, and another usage, blobscription. 
\autoref{fig:non_efficient} presents a diagram of these inefficient strategies. 
A Continuous Sending pattern is when the type-3 transaction sender sends more than one transaction in succession that are ultimately contained within the same block. These blobs can be put into a single type-3 transaction instead of multiple ones, which would require duplicate payments. 
Double Sending occurs when a rollup transmits the identical blob multiple times. A type-3 transaction that carries a second identical blob is invalid because the first successful transaction completes the validation of the blob data. These two strategies account for 7\% and 10\% of the total transactions, respectively. Additionally, it is important to mention that blobscription is observed in 106{,}051 transactions in the network, representing 8\% of the total. This section evaluates the impact and economic loss to the network of the inefficient strategies, showing the loss of 186.92 ETH and an average delay in network inclusion of 19 seconds.

\parhead{Economic Cost.~}
We calculate the transaction fee (including base fee and priority fee) for duplicate transactions in Continuous Sending and Double Sending. ~\autoref{fig:cs_loss} plots the cumulative losses from the Double Sending and Continuous Sending strategies for rollups after EIP-4844. 
The left y-axis shows the cumulative loss for these two strategies, and the right y-axis shows the total number of type-3 transactions. We use light blue bars to show the most rollups per month using these two strategies, and the yellow line represents cumulative losses. Since EIP-4844, rollups have lost 186.92 ETH. while Scroll sends the largest volume of transactions using the two inefficient strategies, it is worth noting that the number of these strategies has been decreasing since May, with less than 1 transaction in August. Another finding is that the use of the continuous sending strategy almost stopped in April, this is because Linea sent significantly fewer type-3 transactions in April.

\parhead{Network Delay.~}
The presence of a large number of blobscription transactions violates the original intent of the EIP-4844 design and interferes with the normal rollup validation process. Since a block can only carry a maximum of 6 blobs, the massive number of blobscription transactions in the public mempool may cause a delay in the normal type-3 transactions. 
We further measure the impact of blobscription on network latency. We analyze 100{,}504 blobscription transaction during a period of burst of blobscription usage. During this period, there were 151{,}688 type-3 transaction sent from the rollups, and 21{,}282 of them were affected by blobscription. The results show that the average latency of the affected type-3 transaction was 43 seconds, while the other type-3 transactions only had to wait 24 seconds. The presence of blobscription in the network introduces a delay of nearly 19 seconds to the rollups.
Our finding exposes that if a large number of blobs that are not used for data validation are consistently present within the network, this causes a huge impact on the process of uploading type-3 transactions for rollups, which in turn affects the efficiency of the rollup application.

\end{document}